\documentclass[sigconf, nonacm]{acmart}

\usepackage{booktabs}
\usepackage{multirow}
\usepackage{graphicx}
\usepackage{array}
\usepackage{subfigure}
\usepackage{tabularx}

\begin{document}

\title{QP-OneModel: A Unified Generative LLM for Multi-Task Query Understanding in Xiaohongshu Search}

\author{
  \textbf{Jianzhao Huang$^*$, 
  Xiaorui Huang$^*$, 
  Fei Zhao, 
  Yunpeng Liu, 
  Hui Zhang, 
  Fangcheng Shi,
  Congfeng Li, 
  Zechen Sun, 
  Yi Wu, 
  Yao Hu, 
  Yunhan Bai$^\dagger$, 
  Shaosheng Cao$^\dagger$} \\
  Xiaohongshu Inc.,  China \\
  \texttt{caoshaosheng@xiaohongshu.com}
}

\thanks{$^*$Equal contribution.}
\thanks{$^\dagger$Corresponding authors.}

\renewcommand{\shortauthors}{Huang et al.}

\begin{abstract}
Query Processing (QP) serves as a critical bridge between user intent and content supply in large-scale search engines for Social Network Services (SNS). Traditional QP systems typically rely on pipelines of isolated discriminative models (e.g., BERT), which suffer from limited semantic understanding capabilities and high maintenance overhead.
While Large Language Models (LLMs) offer a potential solution, existing approaches largely remain fragmented by optimizing sub-tasks in isolation, which neglects the intrinsic semantic synergy across sub-tasks and necessitates independent model iterations. Moreover, standard generative methods often lack grounding in SNS scenarios, failing to bridge the distributional gap between open-domain corpora and the informal linguistic patterns of SNS data, while also struggling to adhere to the rigorous business definitions of specific QP tasks.
In this work, we present QP-OneModel, a Unified Generative LLM for Multi-Task Query Understanding in the SNS domain. Specifically, we reformulate heterogeneous query processing sub-tasks into a unified sequence generation paradigm, while adopting a progressive three-stage alignment strategy culminating in multi-reward Reinforcement Learning to ensure robust adaptation. Furthermore, QP-OneModel generates intent descriptions as a novel high-fidelity semantic signal unavailable in traditional pipelines, effectively augmenting downstream tasks such as query rewriting and ranking.
Extensive offline evaluations show that QP-OneModel achieves a 7.35\% gain in overall score over discriminative baselines, with significant F1 boosts in complex NER (+9.01\%) and Term Weighting (+9.31\%). It also exhibits superior generalization on unseen tasks, surpassing the 32B-parameter model by 7.60\% in accuracy. QP-OneModel has been fully deployed to serve the Search Results Page traffic at Xiaohongshu. Online A/B tests confirm its industrial value, optimizing retrieval relevance (DCG) by 0.21\% and lifting user retention by 0.044\%.
\end{abstract}

\maketitle

\section{Introduction}
Driven by the exponential growth of the mobile Internet, Social Network Services (SNS) have evolved into core infrastructure for social interaction and information dissemination~\cite{xia2013socially}. Within this ecosystem, Query Processing (QP) serves as the fundamental upstream module responsible for decoding user intent and bridging the gap between raw queries and content supply, thereby directly dictating retrieval relevance and overall user satisfaction~\cite{croft2010search,huang2013dssm,li2019deep}.

To accurately interpret vague or colloquial user queries, industrial QP systems rely on a series of fundamental sub-tasks, including Named Entity Recognition (NER), Word Segmentation, Term Weighting, and Query Taxonomy~\cite{nadeau2007survey,lample2016neural,zheng2013deep,robertson2009probabilistic,dai2019deepct}. Traditional QP systems typically implement these tasks via a pipeline of isolated discriminative models (e.g., BERT-based sequence taggers and classifiers)~\cite{devlin2018bert,vaswani2017attention}, which suffer from two fundamental limitations: First, they possess limited semantic understanding capabilities, often struggling to capture complex or nuanced user intents, particularly when confronting long-tail queries or data distribution shifts~\cite{blitzer2007biographies,eisenstein2013bad,bamman2014distributed}. Second, they incur high maintenance overhead, as even minor updates to classification taxonomies or input schemas necessitate retraining the entire model, significantly hindering rapid iteration in fast-paced industrial environments~\cite{sculley2015hidden}.

The emergence of Large Language Models (LLMs) offers a potential solution~\cite{brown2020language,touvron2023llama,bai2023qwen,openai2023gpt4}, yet effective deployment in industrial SNS QP scenarios faces critical challenges~\cite{metzler2021rethinking}. 
First, existing generative approaches remain fragmented, typically optimizing sub-tasks in isolation and neglecting the intrinsic semantic synergy across tasks.~\cite{raffel2020t5,wei2022chain} 
Second, general LLMs struggle with the distinct distributional shift of SNS data, failing to capture the extreme sparsity and rapid lexical evolution (e.g., emerging slang) inherent in social media~\cite{eisenstein2013bad,han2013lexical,bamman2014distributed}. 
Third, the complexity of business rules coupled with data scarcity renders standard fine-tuning inadequate~\cite{ouyang2022training,zhao2023survey}. Industrial QP is governed by fine-grained and often counter-intuitive business definitions~\cite{sculley2015hidden}. Constrained by the limited volume of up-to-date gold-standard data, standard supervised fine-tuning (SFT) often devolves into rote memorization of surface patterns rather than truly internalizing these complex logical boundaries~\cite{zhao2023survey,ouyang2022training}.

In this work, we propose QP-OneModel, which integrates unified generative modeling of heterogeneous QP tasks, a robust domain-specific backbone, and a three-stage progressive alignment strategy. Instead of relying on isolated discriminative heads, our model generates a structured sequence containing all analytical results in a single pass, utilizing global query context to resolve local ambiguities. To mitigate the distributional shift between general corpora and the SNS domain, we leverage the RedOne~\cite{zhao2025redone, zhao2025redone2.0} backbone. By harnessing its innate grasp of informal linguistic patterns and emerging slang, QP-OneModel effectively bridges the gap between raw user queries and structured representations.

To overcome the challenges of complex business rules and data scarcity, we design a progressive three-stage alignment strategy that bootstraps the model from broad knowledge to precise logic internalization. 
1) First, we perform \emph{Knowledge Injection} via a mixed-training strategy, integrating massive auxiliary datasets derived from historical logs to establish a broad semantic foundation. 
2) Subsequently, we transition to \emph{Target Distribution Alignment}, utilizing strictly real-time human annotations to eliminate residual noise and align with evolving trends. 
3) Finally, to prevent the rote memorization typical of standard SFT, we introduce a \emph{Multi-Reward Reinforcement Learning (RL)} stage. By optimizing a composite reward function that balances all sub-tasks, this process drives the model to internalize underlying reasoning logic, further improving performance on tasks demanding deeper semantic understanding, such as NER and Term Weighting.

Furthermore, leveraging the unified generative paradigm, QP-OneModel produces Intent Descriptions, a natural language narrative of the user's search goal derived from the joint modeling of QP sub-tasks. This novel output serves as a high-fidelity semantic signal, effectively augmenting downstream applications such as query rewriting and ranking.

In summary, our main contributions are as follows:
\begin{itemize}
\item Unified Generative Paradigm for QP: We formulate heterogeneous QP sub-tasks into a single sequence-to-sequence generation task. This unified architecture breaks the dependencies of cascaded pipelines, enabling the global optimization of intrinsic task correlations.
\item Progressive Alignment Strategy: We employ a progressive three-stage alignment strategy, alongside an SNS-specific backbone. This systematically addresses the challenges of complex business rules and data scarcity, guiding the model from general understanding to the rigorous internalization of business rules.
\item High-Fidelity Semantic Augmentation: We introduce generative Intent Descriptions as a novel output. Enabled by the unified modeling of structural and semantic information, these descriptions provide downstream systems with distinct, high-quality intent signals inaccessible to discriminative models.
\item Industrial-Scale Validation: We provide extensive empirical evidence from both offline evaluations and large-scale online A/B testing at Xiaohongshu. Offline results demonstrate that QP-OneModel achieves a 7.35\% overall improvement over BERT-based baselines, and surpasses the significantly larger Qwen3-32B by 7.60\% accuracy on the unseen Document Intent task. Online deployment also shows substantial gains, where deploying fundamental signals optimizes retrieval relevance (DCG) by 0.21\%, and leveraging generated intent descriptions increases user retention by 0.044\%.
\end{itemize}

\begin{figure*}[t]
    \centering
    \includegraphics[width=.98\textwidth]{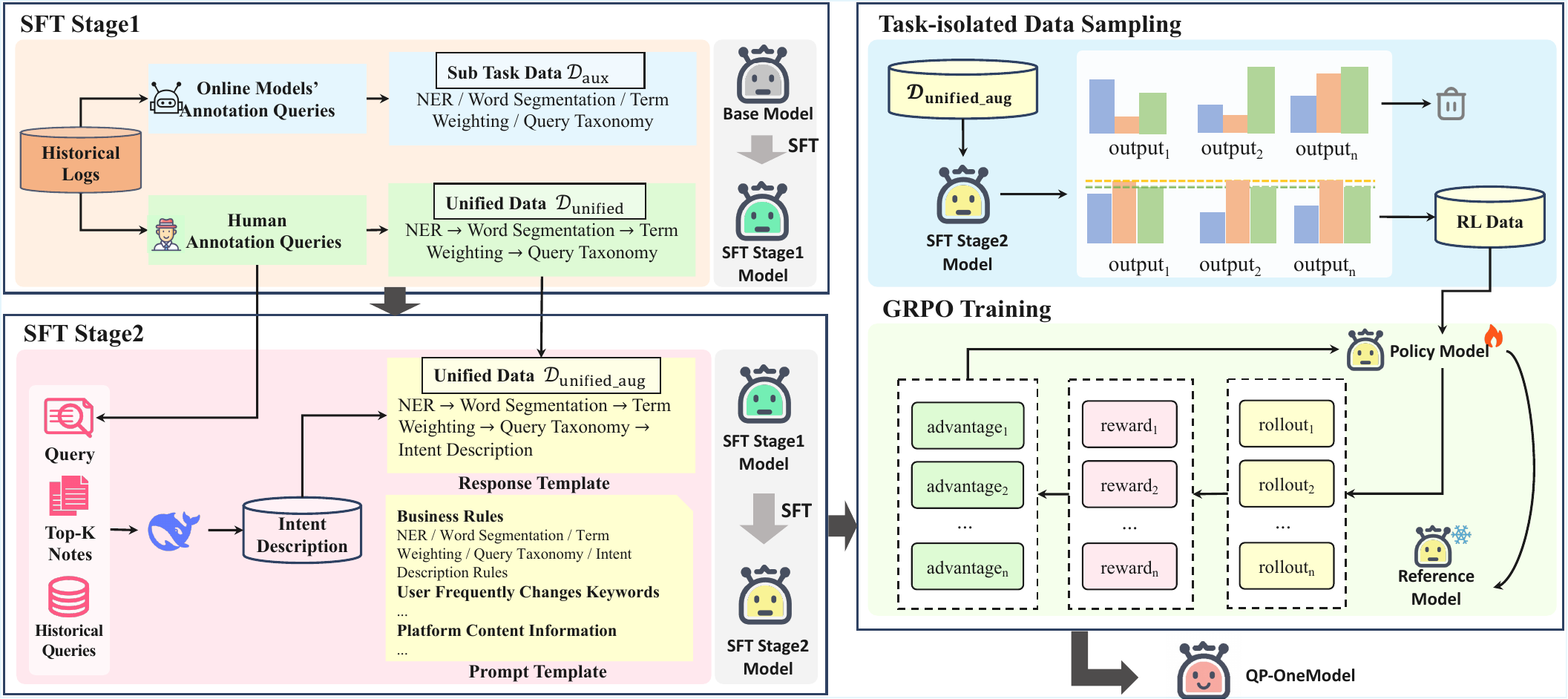}
    \caption{The overall framework of QP-OneModel, covering data construction, multi-stage SFT, and reinforcement learning.}
    \label{fig:qp_overview}
\end{figure*}
\section{Related Work}

\subsection{Query Processing in Industrial Search}
Query Processing (QP) bridges raw user queries and downstream retrieval by producing structured semantic signals, including NER, segmentation, term weighting, and taxonomy/intent labels~\cite{croft2010search,nadeau2007survey,robertson2009probabilistic,dai2019deepct}. 
Early industrial solutions relied on statistical modeling and structured prediction, with CRFs as a representative paradigm for sequence labeling tasks~\cite{lafferty2001conditional}. 
More recently, Transformer-based PLMs have become the dominant backbone: QP sub-tasks are typically cast as sequence labeling or classification and solved by discriminative encoders such as BERT~\cite{vaswani2017attention,devlin2018bert}. 
Despite their strong performance over earlier statistical methods, production QP is commonly implemented as cascaded pipelines of independently trained components, which introduce non-trivial system overhead and error propagation~\cite{sculley2015hidden}. 
Such pipelines are also sensitive to domain shift and long-tail traffic, a challenge amplified in social media where vocabulary and entities evolve rapidly~\cite{blitzer2007biographies,eisenstein2013bad,bamman2014distributed,han2013lexical}.

\subsection{LLMs for Information Retrieval}
LLMs have been increasingly adopted in IR for query rewriting, query expansion, relevance modeling, and generative retrieval~\cite{metzler2021rethinking,jagerman2023query,wang2023query}. 
In particular, prompting LLMs to generate expansions or pseudo-documents has shown effectiveness in mitigating lexical mismatch between queries and documents~\cite{jagerman2023query,wang2023query}. 
Recent vertical-search studies also report substantial gains in relevance estimation by leveraging LLM reasoning capabilities~\cite{dong2025taosr1}. 
Nevertheless, industrial QP remains largely dominated by discriminative PLM pipelines, and applying LLMs to it is still comparatively under-explored~\cite{devlin2018bert,sculley2015hidden}.

\subsection{Unified Generative Modeling and Alignment}
A natural direction to reduce pipeline fragmentation is to unify heterogeneous tasks within a single model. 
Unified text-to-text learning shows that diverse NLP tasks can be cast as sequence-to-sequence generation, enabling a unified interface and joint optimization~\cite{raffel2020t5}. 
Recently, industrial studies have explored end-to-end generative unification in search applications, including unified frameworks for query suggestion and search modeling~\cite{guo2025onesug,chen2025onesearch}. 
These trends are consistent with scaling-law observations that larger generative models improve transfer and robustness given appropriate training signals~\cite{kaplan2020scaling,brown2020language}.
Deploying unified generative models in production also requires alignment to enforce platform-specific constraints. 
The standard recipe combines supervised fine-tuning with reinforcement learning from human feedback, often optimized using PPO~\cite{ouyang2022training,schulman2017proximal}.
For settings where correctness can be directly measured, reinforcement learning with verifiable rewards provides an alternative that optimizes explicit outcomes rather than a learned preference model~\cite{shao2024deepseekmath,guo2025deepseekr1}. 
Motivated by these advances, our work unifies SNS QP sub-tasks as structured generation and adopts progressive alignment to improve robustness under evolving language and rule-intensive definitions.

\section{Preliminaries}
\label{sec:preliminaries}

Query Processing (QP) serves as the foundational layer of the search engine, transforming raw user inputs into structured semantic representations to facilitate effective information retrieval. Formally, given a query sequence $q$, we define the following four core sub-tasks employed in our industrial environment and introduce a new generative sub-task:

\noindent \textbf{Named Entity Recognition (NER).} 
NER aims to identify semantic spans within $q$ and classify them into a predefined ontology.

\noindent \textbf{Word Segmentation.} 
For languages lacking explicit delimiters, Word Segmentation partitions the continuous character sequence $q$ into discrete lexical units (terms).

\noindent \textbf{Term Weighting.} 
Term Weighting quantifies the semantic contribution of each segmented term to the user's core intent. We formulate this as a token-level classification task using a discrete 4-tier relevance scale $s \in \{0, 1, 2, 3\}$. The hierarchy ranges from \textit{Level 0} (functional stop words) to \textit{Level 3} (core intent carriers), directly governing the mandatory matching logic of the search engine.

\noindent \textbf{Query Taxonomy.} 
Query Taxonomy associates a query $q$ with multiple relevant categories to facilitate vertical-specific retrieval strategies. We formulate this as a multi-label classification task that yields a ranked sequence of labels. The label order encodes relevance, prioritizing the Top-1 label as the dominant intent.

\noindent \textbf{Intent Description.} This task generates a natural language narrative that explicates the user's underlying search goal.

\section{Methodology}
In this section, we present the methodological framework for QP-OneModel, addressing the twin challenges of error propagation in fragmented pipelines and the misalignment of general LLMs with specific search business logic in SNS. Section~\ref{sec:qptask} establishes the foundation of our approach by reformulating the entire QP workflow as a unified sequence generation problem, enabling end-to-end optimization of inter-task dependencies.  Section~\ref{sec:qpprompt} details our structured prompting interface that integrates configurable business rules with real-time contextual signals. Section~\ref{sec:alignment} presents a Progressive Three-Stage Alignment Strategy, which systematically bootstraps the model from broad domain knowledge injection (Section~\ref{sec:stage1}) to precise target distribution alignment (Section~\ref{sec:stage2}), and culminates in the internalization of complex business logic via Multi-Reward Reinforcement Learning (Section~\ref{sec:stage3}).

\subsection{Query Processing as Unified Sequence Generation}
\label{sec:qptask}

We reformulate the entire QP workflow as a unified sequence-to-sequence generation task.
Formally, given an input query $q$, static task instruction $I$, configurable business rules $R$, and dynamic contextual information $C$ (e.g., user rewrite history, candidate notes), our model $\pi_\theta$ generates a structured output $\mathbf{y} = (y_1, y_2, \ldots, y_T)$ in an autoregressive manner:
\begin{equation}
\pi_\theta(\mathbf{y} \mid I, R, C, q) = \prod_{t=1}^T \pi_\theta(y_t \mid I, R, C, q, y_{<t}),
\end{equation}
where $\pi_\theta(y_t \mid \cdot)$ represents the probability of generating the $t$-th token conditioned on the complete input configuration and all previously generated tokens $y_{<t}$.

The output $\mathbf{y}$ is structured as a JSON object containing results for all QP sub-tasks:
\begin{equation}
\mathbf{y} = \{ \text{entities}, \text{segments}, \text{weights}, \text{category}, \text{intent\_desc} \}.
\end{equation}

The task execution order is deliberately designed: Named Entity Recognition → Word Segmentation → Term Weighting → Query Taxonomy → Intent Description.
This sequential dependency allows downstream sub-tasks to leverage upstream results, thereby maximizing task synergy.
For instance, accurate entity recognition directly refines word segmentation, and both contribute to optimizing term weighting.

Given a training dataset $\mathcal{D} = \{(I, R, C^{(i)}, q^{(i)}, \mathbf{y}^{(i)})\}_{i=1}^N$, where $\mathbf{y}^{(i)}$ represents the ground-truth structured output, our objective is to maximize the likelihood of generating correct outputs:
\begin{equation}
\max_{\pi_\theta} \mathbb{E}_{(I, R, C, q, \mathbf{y}) \sim \mathcal{D}}
\left[ \log \pi_\theta(\mathbf{y} \mid I, R, C, q) \right].
\end{equation}

This unified generative paradigm eliminates the need for constructing task-specific architectures or maintaining dedicated model instances, and enables end-to-end optimization, naturally capturing inter-task dependencies that remain inaccessible to isolated single-task models.

\subsection{Business-Aware Prompt Design}
\label{sec:qpprompt}

\begin{figure}[t]
    \centering
    \includegraphics[width=\columnwidth]{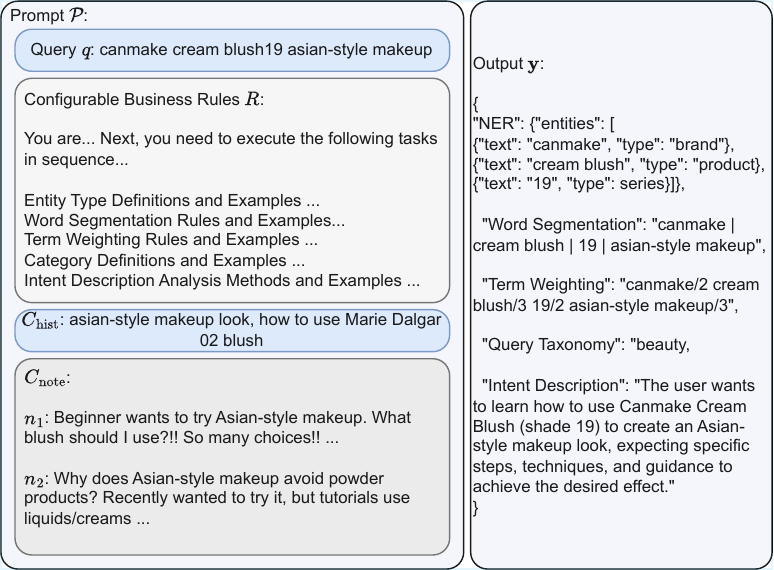}
    \caption{Illustration of the Business-Aware Prompt schema. The prompt $\mathcal{P}$ integrates configurable business rules $R$ with dynamic contexts, including user rewrite history $C_{\text{hist}}$ and candidate notes $C_{\text{note}}$. This context-rich formulation guides the model to generate a unified JSON output covering all QP sub-tasks.}
    \label{fig:qp_prompt}
\end{figure}

As illustrated in Figure~\ref{fig:qp_prompt}, our business-aware prompt integrates three critical components:

\textbf{1) Configurable Business Rules.}
We encode fine-grained operational knowledge as explicit instructions.

This design allows operations staff to rapidly update the prompt to handle emerging scenarios or enforce new policies without the need for retraining. These rules include:
\begin{itemize}
    \item \emph{Entity Definitions}: Criteria for identifying entities (e.g., brands, KOLs, trending IPs).
    \item \emph{Segmentation Rules}: Custom segmentation rules that handle terms such as SNS slang and emerging expressions.
    \item \emph{Term Weighting Rules}: Logic for assigning importance scores to query terms.
    \item \emph{Category Taxonomy}: Taxonomy definitions including boundary cases and disambiguation guidelines.
    \item \emph{Intent Analysis}: Methods and examples for analyzing complex user search intents.
\end{itemize}

\textbf{2) User Rewrite History ($C_{\text{hist}}$).}
Search behaviors are rarely isolated and users frequently refine queries sequentially to clarify intent or explore topics. We dynamically inject the user's rewrite history within the current session into the prompt as temporal context:
\begin{equation}
C_{\text{hist}} = \{q_{t-k}, q_{t-k+1}, \ldots, q_{t-1}\},
\end{equation}
where $q_t$ is the current query and the set represents the $k$ most recent historical queries. This context aids in disambiguating vague inputs and capturing evolving user intent throughout the session.

\textbf{3) Candidate Notes ($C_{\text{note}}$).}
For queries that are semantically ambiguous, time-sensitive, or deeply platform-specific, we augment the prompt with candidate notes retrieved from the platform's content pool:
\begin{equation}
C_{\text{note}} = \{n_1, n_2, \ldots, n_m\},
\end{equation}
where $n_j$ denotes a high-relevance note retrieved via preliminary matching. These notes serve as real-time semantic anchors, helping the model ground abstract queries in actual platform content. 

The final composite prompt $\mathcal{P}$ is constructed by concatenating the base task instruction $I$, configurable rules $R$, current query $q$, and dynamic contexts $C = \{C_{\text{hist}}, C_{\text{note}}\}$:
\begin{equation}
\mathcal{P} = I \oplus R \oplus q \oplus C,
\end{equation}
where $\oplus$ denotes concatenation with appropriate delimiters.

\subsection{Progressive Three-Stage Alignment Strategy}
\label{sec:alignment}

Directly fine-tuning a general LLM on limited unified QP data leads to suboptimal generalization due to the distributional gap between pretraining corpora and SNS queries, as well as insufficient coverage of complex business rules~\cite{zhao2025redone,zhao2025redone2.0}.
To systematically bootstrap the model from broad domain knowledge to precise business logic internalization, we propose a progressive three-stage alignment strategy, as illustrated in Figure~\ref{fig:qp_overview}.

\subsubsection{\textbf{Knowledge Injection via Task Decomposition and Mixed-SFT}}
\label{sec:stage1}

The primary bottleneck in training a unified QP model is the scarcity of real-time, human-annotated data covering all sub-tasks jointly.
Manual annotation of unified structured outputs (JSON format) is extremely labor-intensive and costly, resulting in limited training samples ($\sim 10^5$ samples).

To overcome this data scarcity, we employ a \emph{Task Decomposition} strategy to construct auxiliary datasets from massive historical user logs.
Specifically, we leverage the legacy pipeline—comprising separate models for each QP sub-task—to generate pseudo-labels on unlabeled queries sampled from search logs:
\begin{equation}
\mathcal{D}_{\text{aux}} = \{(q, \mathbf{y}_k^{\text{pseudo}})\}_{k \in \mathcal{T}, q \in \mathcal{Q}_{\text{log}}},
\end{equation}
where $\mathcal{T}$ denotes the set of QP sub-tasks, $\mathcal{Q}_{\text{log}}$ represents the large-scale query pool from historical logs, and $\mathbf{y}_k^{\text{pseudo}}$ is the pseudo-labeled output for sub-task $k$ produced by the legacy system.

While these pseudo-labels are noisy and incomplete (each sample covers only a single sub-task), they provide broad coverage of diverse query patterns at scale ($\sim 10^7$ samples).
To mitigate the semantic drift introduced by noisy pseudo-labels, we adopt a \emph{Mixed-Training} strategy that integrates the limited  human-annotated unified data alongside these auxiliary datasets.
Formally, we optimize:
\begin{align}
\mathcal{L}_{\text{Stage1}}(\theta) &= \mathbb{E}_{(I, R, C, q,\mathbf{y}) \sim \mathcal{D}_{\text{unified}}} \left[ -\log \pi_\theta(\mathbf{y} \mid I, R, C, q) \right] \nonumber \\
&\quad + \lambda \sum_{k \in \mathcal{T}} \mathbb{E}_{(q,\mathbf{y}_k) \sim \mathcal{D}_{\text{aux}}^k} \left[ -\log \pi_\theta(\mathbf{y}_k \mid q) \right],
\end{align}
where $\mathcal{D}_{\text{unified}}$ denotes the small-scale human-annotated unified dataset, $\mathcal{D}_{\text{aux}}^k$ represents the auxiliary dataset for sub-task $k$, and $\lambda$ is a hyperparameter balancing the two data sources.

This mixed-training approach ensures that the model aligns with accurate business logic from $\mathcal{D}_{\text{unified}}$ while simultaneously absorbing broad QP knowledge from the large-scale auxiliary data.

\subsubsection{\textbf{Target Distribution Alignment}}
\label{sec:stage2}

Upon establishing a broad semantic foundation in Stage 1, we transition to a fine-tuning stage utilizing exclusively the real-time, human-annotated unified data $\mathcal{D}_{\text{unified}}$.
This phase is designed to bridge the distributional gap between the historical logs used in Stage 1 and the current online environment.

By focusing strictly on the most recent human-annotated samples, we eliminate residual noise from the auxiliary tasks and ensure the model is precisely aligned with:
(1) the rigorous unified schema where all sub-tasks are jointly annotated and mutually consistent, and
(2) the rapidly evolving linguistic trends of social media.

The training objective for Stage 2 is standard supervised fine-tuning:
\begin{equation}
\mathcal{L}_{\text{Stage2}}(\theta) = -\mathbb{E}_{(I, R, C, q, \mathbf{y}) \sim \mathcal{D}_{\text{unified}}} \left[ \log \pi_\theta(\mathbf{y} \mid I, R, C, q) \right].
\end{equation}

\subsubsection{\textbf{Logic Internalization via Multi-Reward RL}}
\label{sec:stage3}

While supervised fine-tuning effectively teaches the model to mimic reference outputs, it often leads to rote memorization of surface patterns rather than true internalization of underlying business logic. To address this limitation, we introduce a Reinforcement Learning (RL) stage that optimizes the model to maximize task-specific rewards.

\paragraph{\textbf{Reward Design.}}
Given the multi-faceted nature of QP tasks, we design a composite reward function that balances performance across all sub-tasks:
\begin{equation}
\mathcal{R}(\hat{\mathbf{y}}, \mathbf{y}) = \sum_{k \in \mathcal{T}} w_k \cdot r_k(\hat{\mathbf{y}}_k, \mathbf{y}_k),
\end{equation}
where $\hat{\mathbf{y}}_k$ and $\mathbf{y}_k$ are the predicted and ground-truth outputs for sub-task $k$ respectively, and $w_k$ is the weight reflecting the business importance of task $k$.

To ensure strict consistency between the optimization objective and the final performance evaluation, we define $r_k(\cdot, \cdot)$ directly as the evaluation metric used for that specific sub-task. The detailed definitions of these metrics are provided in Section~\ref{sec:exp_settings}.

\paragraph{\textbf{Group Relative Policy Optimization.}}
To optimize the policy $\pi_\theta$, we employ Group Relative Policy Optimization (GRPO)~\cite{shao2024deepseekmath}, a variant of Proximal Policy Optimization (PPO)~\cite{schulman2017proximal} that has demonstrated strong empirical performance on verifiable tasks.

The GRPO objective is:
\begin{align}
\mathcal{J}_{\mathrm{GRPO}}(\theta) 
&= \mathbb{E}_{(I,R,C,q,\mathbf{y}) \sim \mathcal{D}_{\text{GRPO}}, \{\hat{\mathbf{y}}_i\}_{i=1}^G \sim \pi_\theta} \nonumber \\
&\quad \Bigg[ \frac{1}{G} \sum_{i=1}^G \frac{1}{|\hat{\mathbf{y}}_i|} \sum_{t=1}^{|\hat{\mathbf{y}}_i|} \log \pi_\theta(\hat{y}_{i,t} \mid I, R, C, q,\hat{\mathbf{y}}_{i,<t}) \cdot \hat{A}_i \\
&\quad\quad - \beta \;\mathbb{D}_\mathrm{KL}(\pi_{\theta} \| \pi_{\mathrm{ref}}) \nonumber \Bigg].
\end{align}

\paragraph{\textbf{Data Sampling.}}
As QP-OneModel unifies multiple downstream query tasks, a single rollout of a specific prompt yields rewards derived from multiple sub-tasks simultaneously. Directly applying group-wise normalization to the aggregated reward masks performance discrepancies across sub-tasks, thereby hindering the model from identifying which specific sub-tasks underperform.

To mitigate this issue, we leverage the Stage 2 SFT model to filter the initial dataset, specifically retaining samples that exhibit divergent rewards on a target sub-task while maintaining consistent rewards across all other sub-tasks. The resulting subset constitutes the final training dataset for GRPO, denoted as $\mathcal{D}_{\text{GRPO}}$. This strategy enables precise reward attribution in multi-task settings, thereby enhancing the stability and efficacy of policy gradient updates.

\section{Experiments}

\begin{figure*}[t]
    \centering
    \includegraphics[width=.98\textwidth]{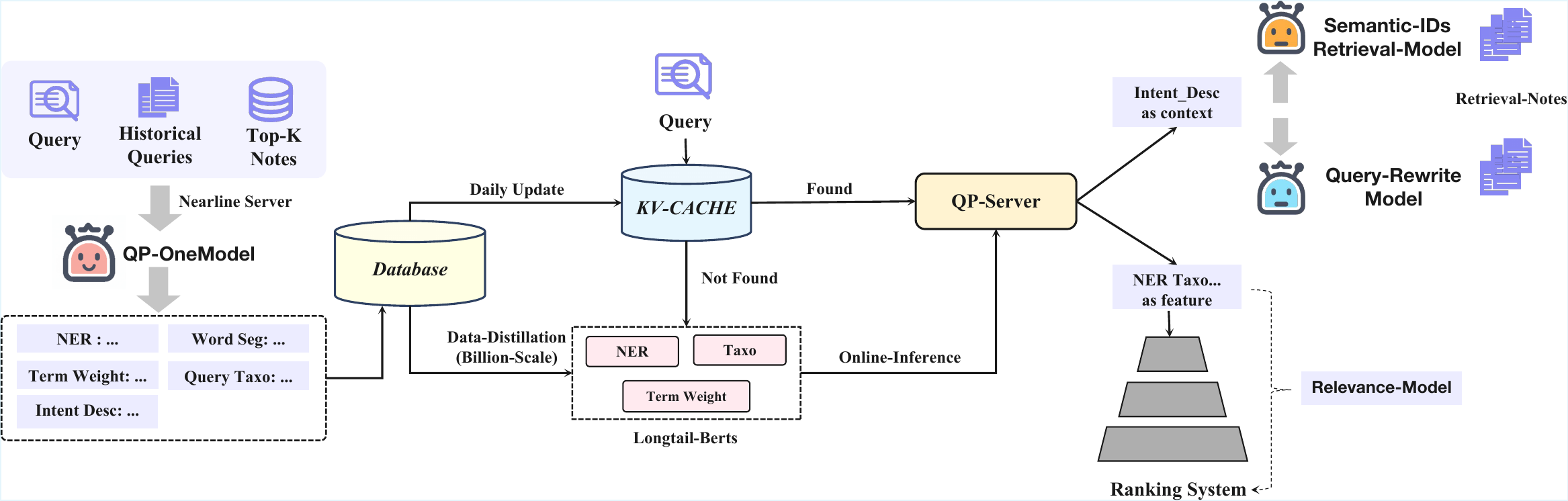}
    \caption{Overview of the deployment architecture. The framework utilizes a nearline inference strategy where QP-OneModel pre-computes results to update the KV-Cache daily. The retrieved structural signals and intent descriptions are then served to downstream tasks such as Query Rewriting and Ranking.}
    \label{fig:deployment}
\end{figure*}

\subsection{Experimental Settings}
\label{sec:exp_settings}

\subsubsection{\textbf{Test Set Construction.}}
To guarantee a robust and unbiased evaluation within a real-world industrial context, we constructed a high-fidelity \textit{Golden Test Set} comprising approximately 2,500 queries randomly sampled from recent production traffic. We adopted a strictly manual annotation pipeline consisting of: (1) Expert-Driven Protocols: Domain experts formulated comprehensive guidelines based on specific domain logic; (2) Specialized Training: Annotators underwent intensive task-specific training; and (3) Iterative Verification: Multi-round cross-validation and expert adjudication were conducted to ensure high inter-annotator agreement. This establishes a reliable benchmark for evaluating QP-OneModel.

\subsubsection{\textbf{Metrics}} 
We employ task-specific metrics to rigorously evaluate model performance across the five sub-tasks.

\noindent \textbf{Standard F1 Tasks.} 
For \textit{Word Segmentation} and \textit{NER}, we adopt the standard F1 Score~\cite{sokolova2009analysis}. 
For NER, we enforce a strict evaluation criterion where a prediction is considered a true positive only if both the entity span boundary and the entity type strictly match the ground truth.

\noindent \textbf{Term Weighting (Joint F1).} 
Term Weighting intrinsically depends on the preceding Word Segmentation, since assigning weights to incorrectly segmented terms is semantically undefined. We define the atomic evaluation unit as a tuple $\langle \text{term}, \text{weight} \rangle$. A predicted unit is counted as correct if and only if both the term boundary and the assigned importance level strictly match the ground truth. We report the F1 Score based on these joint units.

\noindent \textbf{Query Taxonomy.} 
Given that the label order encodes relevance, where the Top-1 label explicitly represents the dominant intent, standard multi-label metrics are insufficient. We design a composite metric averaging the Top-1 Accuracy ($Acc_{top1}$) and the standard F1 Score:
\begin{equation}
    \text{Score}_{QC} = \frac{1}{2} \left( Acc_{top1} + \text{F1} \right)
\end{equation}
This metric balances the precision of the primary intent with the recall of all relevant categories.

\noindent \textbf{Overall.}  The average score of all sub-tasks.

\subsection{Offline Evaluation}

To comprehensively evaluate QP-OneModel, we design experiments to answer the following research questions:
\begin{itemize}
    \item \textbf{RQ1 (Main Results):} How effective is the proposed \textit{QP-OneModel} framework compared to the industry-standard discriminative pipeline?
    \item \textbf{RQ2 (Synergy of Unified Modeling):} Does joint modeling of heterogeneous QP tasks yield positive transfer and synergy?
    \item \textbf{RQ3 (Ablation Study):} How do the domain-specific backbone and the progressive three-stage training strategy contribute to performance?
    \item \textbf{RQ4 (Generalization on Emerging Tasks):} To what extent can the QP-OneModel leverage learned representation to generalize across unseen QP tasks?
\end{itemize}

\subsubsection{\textbf{Main Results: QP-OneModel vs. Discriminative Pipeline (RQ1)}}

We compare QP-OneModel against the online baseline, a pipeline of cascaded BERT-based discriminative models. As shown in Table~\ref{tab:main_results}, QP-OneModel-8B achieves a significant improvement in overall score(+7.35\%) compared to the baseline. Specifically, the baseline struggles with tasks requiring deep semantic understanding and logic, such as NER (74.85\%) and Term Weighting (56.86\%). In contrast, QP-OneModel-8B demonstrates a massive leap in these hard tasks, improving NER F1 by +9.01\% and Term Weighting F1 by +9.31\%. This validates the effectiveness of our comprehensive framework.

Furthermore, it is crucial to note that QP-OneModel-0.6B also substantially outperforms the BERT pipeline across all metrics. It achieves an overall score of 78.37\%, surpassing the baseline by +5.83\%, and delivers remarkable gains in complex tasks, such as improving Term Weighting F1 by +8.59\% (from 56.86\% to 65.45\%). This result highlights that our improvements are derived from the superior effectiveness of the proposed methodology rather than merely scaling up model parameters.

\begin{table*}[t]
  \caption{
        Overall performance comparison on the offline Golden Test Set.
  }
  \label{tab:main_results}
  \centering
  \begin{tabular*}{\textwidth}{l@{\extracolsep{\fill}}cccccc}
            \toprule
            \textbf{Model}  & 
            \textbf{Overall} & 
            \textbf{NER} (F1) &
            \textbf{Word Seg} (F1) &
            \textbf{Term Wgt} (F1) &
            \textbf{Taxonomy} (AVG) \\
            \midrule
            BERT-Pipeline (Baseline) & 72.54 &  74.85 & 85.13 & 56.86 & 73.31 \\
            \midrule
            QP-OneModel-0.6B & 78.81 & 81.64 & 89.65 & 65.45 & 78.5 \\
            QP-OneModel-8B   & \textbf{79.89} &  \textbf{83.86} & \textbf{89.86} & \textbf{66.17} & \textbf{79.68} \\
            \bottomrule
    \end{tabular*}
\end{table*}

\subsubsection{\textbf{Synergy of Unified Modeling (RQ2)}}

A fundamental hypothesis underpinning our methodology is that unifying heterogeneous QP tasks into a single generative paradigm fosters semantic synergy and facilitates positive knowledge transfer through shared latent representations.

To empirically verify this, we conduct a controlled experiment focusing on our Stage 1 phase. We compare the unified model against a Task-Isolated baseline. In this controlled setting, we strictly align all experimental configurations (backbone, loss function, hyperparameters, etc.) with our Stage 1, varying only the data organization. Specifically, we partition the unified multi-task dataset into discrete task-specific subsets, which are then employed to train separate expert models for each QP task independently. Consequently, whereas QP-OneModel operates as a single unified instance, the reported baseline metrics represent an aggregation of the best results achieved by each specialized model on its respective task.

As presented in Table~\ref{tab:synergy}, the Unified model exhibits superior overall score (79.36\% vs. 78.11\%) by significantly outperforming the combined baseline on highly synergistic tasks that exhibit strong mutual dependencies: NER, Word Segmentation, and Term Weighting. This validates the semantic synergy inherent in our sequential generation design.

By modeling the joint probability of all tasks, the unified approach transforms the output of upstream tasks into explicit context for downstream tasks, thereby achieving better performance and efficiency than a suite of isolated models.

\begin{table}[t]
  \caption{Comparison between Unified Modeling and Task-Isolated Training (8B) under the SFT Stage 1 setting.}
  \label{tab:synergy}
  \centering
  \begin{tabular*}{\columnwidth}{l@{\extracolsep{\fill}}ccccc}
        \toprule
        \textbf{Setting} & \textbf{Overall} & \textbf{NER} & \textbf{Seg} & \textbf{TW} & \textbf{Taxo} \\
        \midrule
        Task-Isolated    & 78.11  & 83.02 & 87.98 & 60.92 & \textbf{80.51} \\
        Unified & \textbf{79.36} & \textbf{83.61} & \textbf{89.46} & \textbf{64.78} & 79.59 \\
        \bottomrule
  \end{tabular*}
\end{table}

\subsubsection{\textbf{Ablation Study (RQ3)}}
To understand the source of improvements, we investigate the impact of the backbone model and our training strategies.

\paragraph{\textbf{Impact of Domain-Specific Backbone.}}
To investigate the impact of backbone initialization, we conducted a comparative analysis within the Stage 1  training phase by switching between the general-purpose Qwen3 backbone~\cite{yang2025qwen3} and  RedOne2.0~\cite{zhao2025redone2.0}, a domain-adapted backbone for SNS. As shown in Table~\ref{tab:backbone_comparison}, the RedOne-based model demonstrates a consistent advantage across all metrics, particularly in  Word Segmentation (+0.25\%) and  Taxonomy (+0.69\%). These gains indicate that domain-adaptive pre-training on massive-scale SNS data effectively injects domain linguistic patterns, thereby providing a significantly better initialization for SNS QP tasks compared to a generic foundation model.

\begin{table}[t]
  \caption{Comparison of different backbones (8B) under the SFT Stage 1 setting.}
  \label{tab:backbone_comparison}
  \centering
  \begin{tabular*}{\columnwidth}{l@{\extracolsep{\fill}}ccccc}
        \toprule
        \textbf{Backbone} & \textbf{Overall} &  \textbf{NER} & \textbf{Seg} & \textbf{TW} & \textbf{Taxo} \\
        \midrule
        Qwen3-based   & 79.04  & 83.32 & 89.21 & 64.73 & 78.90 \\
        RedOne2.0-based & \textbf{79.36}  & \textbf{83.61} & \textbf{89.46} & \textbf{64.78} & \textbf{79.59} \\
        \bottomrule
  \end{tabular*}
\end{table}

\paragraph{\textbf{Contribution of Progressive Alignment Stages.}}
Table~\ref{tab:ablation_stages} validates our three-stage alignment strategy. 
\begin{itemize}
    \item \textbf{Stage 2 (Target Alignment)} yields consistent improvements over Stage 1 in overall score and specific sub-tasks. For the 8B model, overall score reaches 79.44\%, with Taxonomy rising from 79.59\% to 79.70\%. A similar trend is observed in the 0.6B model, where overall score moves from 77.95\% to 78.27\%, with NER improving from 80.88\% to 81.91\%. We suppose that fine-tuning on high-quality real-time unified data helps the model better align with the online data distribution and output schema.
    \item \textbf{Stage 3 (RL)} yields noticeable performance improvements, particularly on tasks that require robust semantic understanding. For the 8B model, we observe gains in Term Weighting (+1.26\%) and NER (+0.24\%). This efficacy extends to the 0.6B model, which achieves a notable increase of +1.35\% in Term Weighting (from 64.10\% to 65.45\%). 
    These results suggest that exploration with verifiable rewards enables models to learn nuanced rules that remain elusive under SFT.

\end{itemize}

\begin{table}[t]
  \caption{Ablation study of training stages across different model scales.}
  \label{tab:ablation_stages}
  \centering
  \begin{tabular*}{\columnwidth}{l@{\extracolsep{\fill}}ccccc}
        \toprule
        \textbf{Stage} & \textbf{Overall}  & \textbf{NER} & \textbf{Seg} & \textbf{TW} & \textbf{Taxo} \\
        \midrule
        \multicolumn{6}{l}{\textit{0.6B Model}} \\
        Stage 1        & 77.95  & 80.88 & 88.85 & 63.71 & 78.35 \\
        Stage 2        & 78.27  & \textbf{81.91} & 88.63 & 64.10 & 78.44 \\
        Stage 3 (Ours) & \textbf{78.81} & 81.64 & \textbf{89.65} & \textbf{65.45} & \textbf{78.50} \\
        \midrule
        \multicolumn{6}{l}{\textit{8B Model}} \\
        Stage 1        & 79.36  & 83.61 & 89.46 & 64.78 & 79.59 \\
        Stage 2        & 79.44  & 83.62 & 89.51 & 64.91 & \textbf{79.70} \\
        Stage 3 (Ours) & \textbf{79.89} & \textbf{83.86} & \textbf{89.86} & \textbf{66.17} & 79.68 \\
        \bottomrule
  \end{tabular*}
\end{table}

\subsubsection{\textbf{Generalization on Emerging Tasks (RQ4)}}
\label{sec:generalization}

To assess the model's capacity for transferring learned knowledge to novel problems, we evaluate its generalization capability on two emerging QP tasks that were not part of the original multi-task training curriculum. This serves as a rigorous benchmark for the model's "meta-understanding" of user search intent. Both evaluations are conducted in a few-shot setting. The new tasks are defined as:
\begin{itemize}
    \item \textbf{Document Intent Recognition (Task 1):} This task classifies queries based on whether the user intends to find document-style content (e.g., articles, long-form guides). As Xiaohongshu diversifies its content ecosystem, accurately identifying this intent is crucial for guiding retrieval and ranking strategies, such as allocating specific quotas or applying ranking boosts for document-centric results.
    \item \textbf{Authority Intent Recognition (Task 2):} This task determines if a query seeks authoritative or official information (e.g., from official brand websites, scientific sources). This signal is vital for downstream AI search applications, guiding decisions like triggering external API calls to high-authority search engines or prioritizing the indexing of content from trusted domains.
\end{itemize}

\paragraph{\textbf{Few-Shot In-Context Learning (ICL) Generalization}}
We first evaluate the models' ability to generalize without any gradient updates, relying solely on in-context learning from a few demonstration examples provided in the prompt. We benchmark against general LLMs (Qwen series)~\cite{qwen2025qwen25,yang2025qwen3} and the domain-specific RedOne family~\cite{zhao2025redone,zhao2025redone2.0}. 

As shown in Table~\ref{tab:new_task_exp}, QP-OneModel-8B demonstrates remarkable ICL capabilities. For \textit{Document Intent}, it achieves an accuracy of 82.40\%, significantly surpassing not only its domain-adapted counterpart, RedOne-8B (75.80\%), but also the much larger Qwen3-32B model (74.80\%). For \textit{Authority Intent}, our model attains a competitive accuracy of 71.56\%, performing on par with the 32B model and outperforming other models in its size class. These results strongly suggest that the multi-task, multi-stage alignment strategy has endowed QP-OneModel with a robust, abstract understanding of query semantics, enabling it to effectively solve unseen tasks through prompting alone.

\paragraph{\textbf{Few-Shot Fine-Tuning Performance}}
Next, we assess the models' adaptability by fine-tuning them on a small set of task-specific data using LoRA~\cite{hu2021lora}. QP-OneModel continues to exhibit superior performance.
For \textit{Document Intent}, QP-OneModel-8B achieves a Macro F1 of 78.60, outperforming the RedOne-8B baseline. 
For \textit{Authority Intent}, QP-OneModel-8B achieves the highest overall score (77.34\%) and Macro F1 (72.76\%) among all evaluated models.
These suggest that the comprehensive training on core QP tasks, such as Term Weighting and NER, enables effective generalization to new domains upon fine-tuning.

\begin{table*}[t]
  \caption{
        Overall performance comparison on Document Intent and Authority Intent tasks. 
        Best and second-best results are marked in \textbf{bold} and \underline{underlined}, respectively.
  }
  \label{tab:new_task_exp}
  \centering
  \begin{tabular*}{\textwidth}{l@{\extracolsep{\fill}}ccccccccc}
        \toprule
        \multirow{3}{*}{\textbf{Model}} & 
        \multirow{3}{*}{\textbf{Size}} & 
        \multicolumn{4}{c}{\textbf{In-Context Learning (Few-shot)}} & 
        \multicolumn{4}{c}{\textbf{Supervised Fine-Tuning (SFT)}} \\
        \cmidrule(lr){3-6} \cmidrule(lr){7-10}
         & & 
        \multicolumn{2}{c}{\textbf{Doc. Intent}} & 
        \multicolumn{2}{c}{\textbf{Auth. Intent}} & 
        \multicolumn{2}{c}{\textbf{Doc. Intent}} & 
        \multicolumn{2}{c}{\textbf{Auth. Intent}} \\
        \cmidrule(lr){3-4} \cmidrule(lr){5-6} \cmidrule(lr){7-8} \cmidrule(lr){9-10}
         & & Acc. & F1 & Acc. & F1 & Acc. & F1 & Acc. & F1 \\
        \midrule
        Qwen2.5-7B-Instruct & 7B  & 51.41 & 42.82 & 34.50 & 28.10 & 91.00 & 75.72 & 76.46 & 71.41 \\
        RedOne-7B           & 7B  & 70.58 & 56.26 & 48.78 & 39.96 & 91.13 & 77.22 & 76.76 & 71.74 \\
        \midrule
        Qwen3-8B            & 8B  & 74.85 & 61.84 & 59.97 & 49.48 & 91.27 & 76.56 & 76.80 & 71.94 \\
        RedOne2.0-8B           & 8B  & \underline{75.80} & \underline{62.58} & 70.56 & \underline{63.00} & \underline{91.80} & 77.88 & 76.36 & 71.11 \\
        \textbf{QP-OneModel-8B (Ours)} & \textbf{8B} & \textbf{82.40} & \textbf{62.99} & \underline{71.56} & 60.41 & \underline{91.80} & \underline{78.60} & \textbf{77.34} & \textbf{72.76} \\
        \midrule
        Qwen3-32B            & 32B & 74.80 & 61.96 & \textbf{71.62} & \textbf{64.16} & \textbf{92.33} & \textbf{79.20} & \underline{77.14} & \underline{72.02} \\
        \bottomrule
  \end{tabular*}
\end{table*}

\subsection{Online Evaluation}
We deployed QP-OneModel online at Xiaohongshu, with the deployment pipeline shown in Figure \ref{fig:deployment}, and utilized the Xiaohongshu A/B testing platform to evaluate its performance. We randomly sampled \textbf{5\%} of the online traffic as the treatment group and another \textbf{5\%} as the control group. To mitigate the impact of traffic fluctuations, the experiment was conducted over a period of at least 14 days. We designed two sets of experiments to evaluate the fundamental signals produced by QP-OneModel and the extrinsic utility of the generated intent descriptions.

\subsubsection{\textbf{Experimental Setup}}

\paragraph{\textbf{Experiment 1: Fundamental Signal Evaluation}}
The goal of this experiment is to assess the accuracy of the core query processing signals. In the treatment group, we replaced the signals (including NER, taxonomy, and term weights) generated by the previous BERT-based model with those produced by QP-OneModel.

\paragraph{\textbf{Experiment 2: Downstream Application (Query Rewriting)}}
This experiment evaluates the effectiveness of the "Intent Descriptions" generated by QP-OneModel in downstream tasks. We utilized \textit{Query Rewriting} as the testbed, which leverages the original query and global statistical signals to generate rewrites that better satisfy the user intent.
\begin{itemize}
    \item \textbf{Control Group}: Uses the current production baseline, a LLM-based query rewriter.
    \item \textbf{Treatment Group}: Augments the baseline model with \textbf{Intent Descriptions}. These descriptions serve as Chain-of-Thought (CoT) supervision during training and RAG inputs during inference.
\end{itemize}

\subsubsection{\textbf{Evaluation Metrics}}

We adopted the following metrics to measure performance. Note that for our platform, small absolute changes in these metrics can indicate statistical significance.

\begin{itemize}
    \item \textbf{DCG 0/1}: A relevance metric where professional annotators label top-8 results as relevant or irrelevant. This metric tracks the cumulative penalty of irrelevant results; a lower value indicates higher ranking precision. An absolute decrease of $\ge 0.15$ is considered statistically significant.
    \item \textbf{Zero/Few-Result Rate}: The percentage of queries returning extremely few results (typically 1-2 items).
    \item \textbf{Note Effective CTR (NECTR)}: Measures clicks leading to meaningful engagement (e.g., long dwell time, likes). An absolute increase of $\ge 0.1\%$ is considered statistically significant.
    \item \textbf{SAU-Retention}: The next-day retention rate of Search Active Users, serving as a core indicator of user satisfaction.
\end{itemize}

\subsubsection{\textbf{Experimental Results}}

The online A/B test results are presented below. All reported changes represent absolute deltas.

\paragraph{\textbf{Fundamental Signal Performance}}

In Experiment 1, replacing the baseline signals with QP-OneModel resulted in improved relevance. The DCG 0/1 metric decreased by 0.21\%, which exceeds the significance threshold of 0.15\%. We attribute this to the synergy between our unified paradigm, which mitigates error propagation, and the RL-driven alignment that enforces complex business logic. Additionally, the 0.4631\% drop in Zero/Few-Result Rate validates our domain-specific backbone. We believe that by mastering informal slang during knowledge injection, the model bridges semantic gaps in long-tail queries where baselines failed.

\paragraph{\textbf{Downstream Application Performance}}
In Experiment 2, incorporating Intent Descriptions into the rewriting task improved user engagement metrics. The Note Effective CTR (NECTR) increased by +0.17\%, which is statistically significant compared to the 0.1\% threshold. Furthermore, the core SAU-Retention metric increased by +0.044\%. These results suggest that the intent signals provided by QP-OneModel contribute to better semantic understanding and user experience in downstream applications.

\begin{table}[t]
  \centering
  \caption{Online A/B testing results. \textbf{Top:} Evaluation of fundamental structural signals (NER, Term Weighting, etc.). \textbf{Bottom:} Verification of the \textbf{Intent Description's effectiveness}, showing that augmenting downstream Query Rewriting task with this novel generative output improves user retention.}
  \label{table_combined_experiments}
  
  \resizebox{\columnwidth}{!}{
    \begin{tabular}{lcc}
      \toprule
      \multicolumn{3}{l}{\textit{\textbf{Fundamental Signal Evaluation}}} \\
      \cmidrule(lr){1-3}
      Model & $\Delta$DCG 0/1 $\downarrow$ & $\Delta$Zero/Few-Result $\downarrow$ \\
      \midrule
      Baseline & 0     & 0 \\
      Ours     & -0.21\% & -0.4631\% \\
      
      \bottomrule \addlinespace[1ex] 
      
      \multicolumn{3}{l}{\textit{\textbf{Downstream Query Rewriting}}} \\
      \cmidrule(lr){1-3}
      Model & $\Delta$SAU-Retention $\uparrow$ & $\Delta$NECTR $\uparrow$ \\
      \midrule
      Baseline  & 0        & 0 \\
      + Intent Desc.  & +0.044\% & +0.17\% \\
      \bottomrule
    \end{tabular}
  }
\end{table}

\subsection{Qualitative Analysis}

\begin{table*}[t] 
\caption{Qualitative analysis of challenging cases. The results demonstrate QP-OneModel's capability to resolve \textbf{semantic sparsity} (e.g., inferring \textit{Estée Lauder} from ``1c1'') and \textbf{context ambiguity} (e.g., distinguishing Military vs. Gaming contexts).}
\label{tab:qualitative_analysis}
\centering
\small
\renewcommand{\arraystretch}{1.3}
\begin{tabularx}{\textwidth}{>{\raggedright\arraybackslash}p{3.5cm} >{\raggedright\arraybackslash}p{4.5cm} X}
\toprule
\textbf{Query Case} & \textbf{Structural Output} \newline (NER \& Taxonomy) & \textbf{Generated Intent Description} \\ 
\midrule
\textbf{1c1} & 
\textit{Beauty / Makeup} \newline \texttt{[1c1: Product Series]} & 
User is selecting \textbf{Estée Lauder foundation}; seeks actual color performance of shade \textit{1c1}, comparison with other shades (e.g., \textit{2c0}), and skin tone suitability advice. \\ 
\midrule
\textbf{163 Guards Tank Regiment} & 
\textit{1. Social Science / Military} \newline \textit{2. Gaming / Console} & 
User seeks to understand the historical background, organizational structure, and combat performance of the \textit{163rd Guards Tank Regiment} (or in-game vehicle stats). \\ 
\midrule
\textbf{retro-x} & 
\textit{Clothing - Outdoor} \newline \texttt{[retro-x: Product Series]} & 
User considers purchasing \textbf{Patagonia Retro-X} fleece; seeks detailed performance evaluations (wind resistance), target audience (kids/adults), and price advice. \\ 
\bottomrule
\end{tabularx}
\end{table*}

Table \ref{tab:qualitative_analysis} presents a qualitative analysis of representative cases, illustrating how QP-OneModel utilizes its generative capabilities to handle challenging queries.

\paragraph{\textbf{Overcoming Semantic Sparsity via Latent Linking}}
For the query \textbf{"1c1"}, which consists solely of an alphanumeric code, discriminative baselines often struggle to extract meaningful features. In contrast, QP-OneModel successfully identifies the unmentioned parent brand (\textit{Estée Lauder}) and infers the user's implicit goal of shade comparison. This highlights the model's ability to leverage internal knowledge to bridge the semantic gap in cryptic inputs.

\paragraph{\textbf{Resolving Polysemy and Enhancing Granularity}}
The model also effectively handles ambiguity and intent deepening. For \textbf{"163 Guards Tank Regiment"}, it captures the multi-faceted nature of the query by identifying both the explicit military fact and the latent gaming context (simulation games). Furthermore, for \textbf{"retro-x"}, the model expands the product entity into actionable attributes such as wind resistance and price, providing downstream rankers with rich semantic signals.

\section{Conclusion}
In this paper, we introduce QP-OneModel, a unified generative framework that reformulates heterogeneous query processing sub-tasks into a single sequence-to-sequence paradigm.
By leveraging the domain-specific RedOne backbone and a progressive multi-reward reinforcement learning strategy, our approach significantly outperforms discriminative baselines, achieving a 7.35\% increase in overall score and boosting F1 scores in complex NER and Term Weighting tasks by 9.01\% and 9.31\%, respectively.
Moreover, it demonstrates robust generalization on unseen QP tasks, surpassing the significantly larger Qwen3-32B by 7.60\% accuracy on Document Intent under ICL setting.
Extensive online A/B tests confirm its industrial value: deploying QP-OneModel's fundamental signals improves DCG 0/1 by 0.21\%, while leveraging its generated intent descriptions for downstream query rewriting yields a 0.044\% uplift in user retention.
Currently, QP-OneModel has been fully deployed to serve the Search Results Page traffic at Xiaohongshu.

\bibliographystyle{ACM-Reference-Format}
\bibliography{reference}

\appendix

\end{document}